\documentstyle[aps,12pt]{revtex}

\begin{document}
\title{Numerical Simulation on Tunnel Splitting of Bose-Einstein Condensate in
Multi-Well Potentials}
\author{Yajiang Hao$^1$, J.-Q. Liang$^1$ and Yunbo Zhang$^{1,2,\thanks{%
email: ybzhang@sxu.edu.cn}}$}
\address{$^1$Department of Physics and Institute of Theoretical Physics, Shanxi\\
University, Taiyuan 030006, P. R. China\\
$^2$Laboratory of Optics and Spectroscopy, Department of Physics, University%
\\
of Turku, FIN-20014 Turku, Finland}
\maketitle

\begin{abstract}
The low-energy-level macroscopic wave functions of the Bose-Einstein
condensate(BEC) trapped in a symmetric double-well and a periodic potential
are obtained by solving the Gross-Pitaevskii equation numerically. The
ground state tunnel splitting is evaluated in terms of the even and odd wave
functions corresponding to the global ground and excited states
respectively. We show that the numerical result is in good agreement with
the analytic level splitting obtained by means of the periodic instanton
method.
\end{abstract}

\section{Introduction}

The experimental realization of Bose-Einstein condensation (BEC) in
double-well trap \cite{Andrews,Hinds,Tiecke,Shin1,Shin2} and optical lattice 
\cite{Anderson,Cataliotti,Greiner,Kramer,Paredes} has stimulated active
research into various aspects of quantum tunneling phenomena such as the
Josephson junction \cite{Dalfovo1,Smerzi,Giovanazzi,Salasnich}, atomic
interferometry \cite{Shin1}, the two-wire waveguide \cite{Hinds}, etc.
Dalfovo \cite{Dalfovo1} suggested a Josephson like effect by considering a
confining potential with two wells separated by a barrier. A difference
between the chemical potentials of the atoms in the two traps can be
achieved, for example, by loading a different number of atoms in the traps.
The first experimental evidence \cite{Cataliotti} of the oscillating atom
current was observed instead in an one-dimensional Josephson Junction array
realized with condensates in a laser standing wave, i.e., an optical lattice 
\cite{Jaksch}. The latest techniques of coherently splitting the condensate
by deforming the single optical trap into two wells serve as a model system
for tunneling in the condensate and provide a perfect demonstration of a
trapped-atom interferometer \cite{Shin1,Shin2}.

The coherent tunneling of BEC between double-well traps results in the level
splitting of the macroscopic ground state and hence the macroscopic
coherence, which has been observed in interference experiments \cite
{Andrews,Shin1}. Recently, the energy-band structure and level splitting due
to quantum tunneling in two weakly linked condensates in the ''phase''
representation have been evaluated in terms of the periodic instanton method 
\cite{Li}, which manifests itself as a powerful tool for the calculation of
the tunneling rate and a good approximation for the dilute boson gas \cite
{Liu}. It, nevertheless, is not able to explicitly take into account the
nonlinear interaction term in the Gross-Pitaevskii equation (GPE) describing
the atom-atom collisions in BEC \cite{Zhang}. It is naturally expected that
as the number density of atoms in the condensates increases, the effect of
the nonlinear interaction between atoms would become important. In the
present paper we solve the GPE numerically in order to have a quantitative
evaluation of the energy level splitting of the ground state for BEC
confined in a symmetric double-well trap and an optical lattice and explore
the dependence of the energy level splitting on the chemical potential and
the $s$-wave scattering length between atoms.

Although the numerical solution of GPE has been developed into a standard
procedure, this never prevents us from seeking efficient analytical methods.
The advantage of a nonperturbative method is that it gives not only an good
description of the tunneling phenomena but also a comprehensive physical
understanding in the context of quantum field theory. The periodic instanton
configurations, which have been shown to be a useful tool in several areas
of research such as spin tunneling, bubble nucleation and string theory,
enable also the investigation of the finite temperature behavior of these
systems. In the case of the Bose-Einstein system, it turns out that the
periodic instanton method is reliable in evaluating the tunnel splitting for
BEC trapped in both the double-well and optical lattice in the regime of
experimental values of the chemical potential and scattering length. The
intention of this paper is to quantitatively compare the deviation of the
instanton result from the exact numerical solution and to address the
applicability of the instanton method to actual experimental situations.

We restricted our discussion to a quasi-one-dimensional (Q1D) BEC since it
is mathematically simple, in which the BEC is prepared in optical and
magnetic traps by putting atoms in cylindrical traps long enough that the
one-particle energy-level spacing in the radial direction exceeds the
interatomic-interaction energy, and the atoms can move effectively in the
axial direction \cite{Gorlitz}.

The paper is organized as follows. In Sec. II we give a brief review of the
mean-field analysis for BEC trapped in external potentials. In Sec. III the
energy level splitting is derived in terms of the instanton method based on
the GPE with, however, the nonlinear interaction term included implicitly in
the chemical potential. In Sec. IV we present the numerical procedure for
solving the GPE and evaluation of the ground state level splitting. The
numerical level splitting is compared with the instanton result in Sec. V
and a brief summary is given in Sec. VI.

\section{GPE for one-dimensional Bose gas}

We are interested in the macroscopic quantum tunneling between the
condensates separated by potential barriers and the main concern here is how
the nonlinear interaction between the atoms would affect the level
splitting. We begin with the energy functional for the condensed bosons of
mass $m$ confined in the external potential $V(x)$ given by 
\begin{equation}
E=\int dx\left[ \frac{\hbar ^2}{2m}\left| \frac{d\psi \left( x\right) }{dx}%
\right| ^2+V(x)\left| \psi \left( x\right) \right| ^2+\frac{U_0}2\left| \psi
\left( x\right) \right| ^4\right] 
\end{equation}
where the order parameter of the condensate $\psi \left( x\right) $ are
normalized to the number of atoms in the condensate $\int dx\left| \psi
\left( x\right) \right| ^2=N$ and the 1D effective interaction constant $%
U_0=2\hbar ^2a/ma_{\bot }^2$ \cite{Petrov,Olshanii} characterizes the
nonlinear interaction in the condensate through $s$-wave scattering length $a
$. Here the radial extension of the ground state wavefunction $a_{\bot }=%
\sqrt{\hbar /m\omega _{\bot }}$ is a typical length scale in the transverse
trap with a confinement frequency $\omega _{\bot }$. The first-order
variation of the energy functional leads to the Gross-Pitaevskii equation
(GPE), $H\psi \left( x\right) =\mu \psi \left( x\right) $, with the chemical
potential $\mu =\left\langle \psi \left| H\right| \psi \right\rangle /N$
calculated as the expectation value of the Hamiltonian 
\begin{equation}
H=-\frac{\hbar ^2}{2m}\frac{d^2}{dx^2}+V(x)+U_0\left| \psi \left( x\right)
\right| ^2.
\end{equation}
The transverse confinement frequency $\omega _{\bot }$ should be large
compared with $\mu /\hbar $ so that the condensate is prepared in one
dimension. On the other hand, for a less strong transverse confinement,
atoms will oscillate in all directions which makes the model not exactly
solvable. As can be seen in the last part of the paper, tunneling would be
greatly enhanced in 3D.

Consider two models of the external potential where atoms can tunnel through
the barriers. The double-well trapping potential of the form 
\begin{equation}
V_{dw}(x)=V_0\left( 1-x^2/x_0^2\right) ^2  \label{dw}
\end{equation}
allows us to investigate the interwell coupling which results in the
splitting of the energy level. The potential barrier of depth $V_0\ $between
the two minima $\pm x_0$ is assumed to be large enough so that the overlap
between the wave functions relative to the two traps occurs only in the
classically forbidden region where the interaction can be ignored. The
optical lattice trapping potential 
\begin{equation}
V_{ol}(x)=V_0\cos ^2\left( k_0x\right) ,  \label{ol}
\end{equation}
on the other hand, formed by the standing wave laser beams with wavevector $%
k_0$ \cite{Greiner}, simulates the sine-Gordon potential which is widely
used in quantum field theory as a periodic field model \cite{Liang}. Quantum
tunneling between many wells leads to the formation of energy bands due to
the spatially periodic potential (\ref{ol}). A path integral calculation 
\cite{Liang} was done for these quantum tunneling models both for vacuum and
excited states neglecting, however, the nonlinear interaction. In our
previous work \cite{Zhang}, the nonlinear interaction between the atoms was
included in the finite chemical potential and we realized that tunneling
occurs at the level of chemical potential. Here we will solve the GPE
numerically and compare the numerical results with the analytical ones.

For convenience we rescale the energies and distances in units of $\hbar
\omega _0$ and oscillator lengths $a_0=\sqrt{\hbar /m\omega _0},$ with $%
\omega _0=\sqrt{V^{\prime \prime }(x_b)/m}$ being the frequency of small
oscillations at the bottom of each well $x_b$ in double-well or optical
lattice traps. The wave function is correspondingly rescaled in units of $%
\sqrt{1/a_0}$\ so that it remains normalized to $N$. The GPE thus takes the
following dimensionless form 
\begin{equation}
\left[ -\frac 12\frac{d^2}{dx^2}+V(x)+U_0\left| \psi \left( x\right) \right|
^2\right] \psi \left( x\right) =\mu \psi \left( x\right)   \label{gpe}
\end{equation}
with the potential barrier $V_0$ and chemical potential $\mu $ measured in
units of $\hbar \omega _0$ and the nonlinear interaction parameter becomes $%
U_0=2aa_0/a_{\perp }^2$. Futhermore, we fix the potential parameters as $x_0=%
\sqrt{8V_0}$ and $k_0=1/2\sqrt{V_0}$ in order to leave us with only one
adjustable parameter, i.e., the potential depth $V_0$. The tunnel splitting
depends on these parameters $x_0$ and $k_0$, which are effectively the
separations between the well bottoms, in a similar way as on $V_0$. In the
case of an optical lattice, the depth of the barrier is usually measured in
units of the recoil energy $E_r=\hbar ^2k_0^2/2m$ of the atoms. It is,
however, not difficult to transfer the energy units between this convention
and ours.

\section{The tunnel splitting evaluated with instanton method}

Quantum tunneling between noninteracting particles localized in two adjacent
wells with macroscopic wave functions $\psi _{+}$, $\psi _{-}$ leads to an
effective energy level splitting $\Delta \mu $, which removes the asymptotic
degeneracy. The wave functions of the ground state $\psi _e$ with even
parity and the first excited state $\psi _o$ with odd parity are
superpositions of the localized wave functions $\psi _{+}$ and $\psi _{-}$ 
\begin{equation}
\psi _e=\left( \psi _{+}+\psi _{-}\right) /\sqrt{2}  \label{even}
\end{equation}
\begin{equation}
\psi _o=\left( \psi _{+}-\psi _{-}\right) /\sqrt{2}  \label{odd}
\end{equation}
with energy eigenvalues $\mu \pm \Delta \mu /2$, respectively. When the
interatomic coupling constant $U_0$ vanishes, the problem reduces to the
solution of a linear Schr\"{o}ndinger equation with the Hamiltonian $H=-%
\frac 12\frac{d^2}{dx^2}+V\left( x\right) $. The nonlinear interaction
increases the chemical potential and even for the system at zero
temperature, tunneling occurs at a higher level $\mu $ than the ground
state. The tunnel splitting can be found with the instanton method \cite
{Zhang,Liang} and is generally expressed as 
\begin{equation}
\Delta \mu =\frac{\omega (\mu )}\pi \exp \left[ -S(\mu )\right]   \label{sp}
\end{equation}
where the imaginary time action $S(\mu )$ is calculated through the barrier
region ($\mu <V(x)$) once from turning point $-b$ to turning point $b$: 
\begin{equation}
S(\mu )=\int_{-b}^b\sqrt{2(V(x)-\mu )}dx
\end{equation}
and the frequency $\omega (\mu )$ appearing in the prefactor is the
frequency of the classical periodic oscillations at energy $\mu >V(x)$ in
the classically accessible region with the boundary determined by the
turning points $b$ and $a$ 
\begin{equation}
\omega (\mu )=\frac \pi {\int_b^a\frac{dx}{\sqrt{2(\mu -V(x))}}}.
\end{equation}
The path integral method \cite{Schulman} has been used in the evaluation of
the tunneling rate prefactor and the barrier $V_0$ between two wells is
assumed to be high enough to safely use the WKB wave functions \cite
{Dalfovo1} in the calculation of the transition amplitude. For the potential
in the form of (\ref{dw}) the level splitting reduces to 
\begin{eqnarray}
\Delta \mu  &=&\frac{\sqrt{1+u}}{2{\cal K}\left( k^{\prime }\right) }\exp
\left( -W\right)   \label{spdw} \\
W &=&\frac{16V_0}3\left( 1+u\right) ^{1/2}\left( {\cal E}\left( k\right) -u%
{\cal K}\left( k\right) \right) ,
\end{eqnarray}
where ${\cal K}\left( k\right) $ and ${\cal E}\left( k\right) $ denote the
complete Jacobian elliptic integral of the first and second kinds
respectively. The corresponding parameters are defined as $u=\sqrt{\frac \mu
{V_0}}$, $k^2=\frac{1-u}{1+u}$, and $k^{\prime 2}=1-k^2$. When the nonlinear
interaction vanishes, the dimensionless chemical potential reduces to $\mu
=1/2$ and the above result turns out to be 
\begin{equation}
\Delta \mu =\sqrt{\frac 2\pi }8V_0^{1/2}\exp \left( -\frac{16V_0}3\right) .
\end{equation}
which resembles the case of the vacuum instanton.

Tunneling between many potential wells splits the level further into as many
sublevels as the number of wells. The Bloch theorem tells us that the
eigenvalues of the periodic potential exhibits an energy band structure in
the tight-binding approximation 
\begin{equation}
E(\theta )=\mu +\frac{\Delta \mu }2\cos \left( \theta a\right) 
\end{equation}
with $\Delta \mu $ the band width of the quantum state with energy $\mu $.
Here the Bloch wave vector $\theta $ is confined to the first Brillouin zone 
$\left[ -\pi /a,\pi /a\right] $ of the optical lattice with a lattice
constant $a=\lambda /2$. For the periodic potential of the optical lattice (%
\ref{ol}), the energy band width reads 
\begin{eqnarray}
\Delta \mu  &=&\frac 1{2\sqrt{2}{\cal K}\left( k^{\prime }\right) }\exp
\left( -W\right)  \\
W &=&4\sqrt{2}V_0\left( {\cal E}\left( k\right) -k^{\prime 2}{\cal K}\left(
k\right) \right) 
\end{eqnarray}
with $k^2=1-\mu /V_0$.

\section{Numerical procedure}

There exist various numerical approaches for studying the energy spectrum
and dynamics of BEC trapped in the external potentials \cite{Books}. In the
present paper, we solve the GPE numerically and find the ground- and first
excited-state wave functions $\psi _e\left( x\right) $ , $\psi _o\left(
x\right) $ where the corresponding energy expectation values $\mu _e$ and $%
\mu _o$ are obtained by direct calculation. The level splitting $\Delta \mu
=\mu _e-\mu _o$ is described as a function of parameters $N$ and $U_0$.

We adopt the Gauss-Seidel method to solve eq. (\ref{gpe}) numerically and
hence start from the diffusion equation 
\begin{equation}
\partial _{t}\psi =-H\psi =:\frac{1}{2}\frac{d^{2}\psi }{dx^{2}}-\frac{1}{2}%
\rho ,  \label{df}
\end{equation}
with a diffusion constant of $\frac{1}{2}$ and a source term $\rho $. As $%
t\rightarrow +\infty $, the wave function relaxes to an equilibrium solution
which means that all time derivatives vanish. As a matter of fact the
diffusion equation eq. (\ref{df}) is obtained from the NLSE eq. (\ref{gpe})
with the time being replaced by a negative imaginary time.

We use the following Crank-Nicholson scheme to discretize eq. (\ref{df}) by
using the space step $h$ and time step $\Delta $ 
\begin{eqnarray}
\frac{\psi _i^{n+1}-\psi _i^n}\Delta  &=&\frac 1{2h^2}[\left( \psi
_{i+1}^{n+1}-2\psi _i^{n+1}+\psi _{i-1}^{n+1}\right)   \label{discrete} \\
&&+\left( \psi _{i+1}^n-2\psi _i^n+\psi _{i-1}^n\right) ]  \nonumber \\
&&-\frac 12\left[ V_i\left( x_i\right) +U_0\left| \psi _i^n\right| ^2\right]
\left( \psi _i^{n+1}+\psi _i^n\right)   \nonumber
\end{eqnarray}
where $\psi _i^n=\psi (x_i,t_n)$ denotes the exact solution at $x_i=ih$ and $%
t_n=n\Delta $. The method is stable, unitary, and second-order accurate in
space and time \cite{Recipes,Ruprecht,Adhikari}. In a lattice of $s$ points
eq. (\ref{discrete}) represents a tridiagonal set with open boundary
conditions or a cyclic tridiagonal set with periodic boundary conditions for 
$i=2,3,...,s-1$. For tridiagonal sets, the whole solution can be encoded
very concisely using the procedures of LU decomposition, forward- and back-
substitution while for cyclic tridiagonal sets, the procedure of
Sherman-Morrison Formula is used \cite{Recipes}. For the double-well case we
choose the space step $h=0.01$, time step $\Delta =0.001$ and, $s=2400$. For
the optical lattice, $\Delta =0.02$ and $s=2500$. The values of these
parameters are chosen to satisfy the stability criterion of the
Crank-Nicholson code.

We start from the initial, trial wave functions ($t=0$) given in eqs.(\ref
{even}) and (\ref{odd}) and choose the eigenstates in the non-interacting
limit as our trial wave functions such that $\psi ^{+}$, $\psi ^{-}$ are the
degenerate eigenstates in the left- and right-well with the same energy
eigenvalue. In our procedure, all of the wavefunctions with even parity
finally evolve into the lowest eigenstate with even parity, i.e., the lower
level state for the double well or the bottom of energy band for the optical
lattice. Similarly those states with odd parity evolve into the lowest state
with odd parity, i.e., the upper level state for the double well or the top
of the energy band for the optical lattice. The boundary and normalization
conditions are implemented at each time step. To test the validity of our
code, the numerical wave functions for a stationary GPE in a spherical trap
is compared with the corresponding results given in \cite{Dalfovo2} and the
agreement is perfect.

\section{Numerical result with the nonlinear interaction}

As an example we consider weak-linked condensates of $^{87}$Rb atoms
confined in multi-traps with frequency $\omega _0=100$Hz as in reference 
\cite{Smerzi} and the corresponding oscillator length is $a_0=2.70\times
10^{-4}$ cm. The $s$-wave scattering length is in the range $%
85a_{bohr}<a_{sc}<140a_{bohr}$, where $a_{bohr}$ is the Bohr radius \cite
{Gardner}. In our analysis we use $a_{sc}=100a_{bohr}$. The transverse
confinement frequency $\omega _{\bot }$ is taken to be $2\pi \times 250$Hz.
The corresponding radial extension $a_{\bot }=6.81\times 10^{-4}$ cm and the
interatomic-interaction constant $U_0=0.06$ (in units of $\hbar \omega _0$),
which corresponds to a weak \cite{Salasnich2} nonlinear interaction such
that we could examine its effect on the level splitting. We always measured
energies in units of $\hbar \omega _0$ and lengths in units of the
oscillator length $a_0$, so $V_0$ is all we need.

As a comparison we first of all deal with the ``noninteracting''\ case. When 
$V_0=5$, the analytical tunnel splitting of the instanton approach given in
eq. (\ref{spdw}) is $\Delta \mu =3.74\times 10^{-11}$ and our numerical
result is $\Delta \mu =3.60\times 10^{-11}$. This again proves the validity
of our numerical simulation. The corresponding wave function is shown in
Fig. 1a. In this paper we are mainly interested in the ground state, and for
this purpose the choice of $s=2400$ is seen to be adequate for most of the
calculations. In Fig. 1b we show the profiles of the even (solid line)- and
odd (dotted line) wave functions{\em \ }$\psi _{e,o}\left( x\right) $ for
344 atoms confined in the trap with height $V_0=5$, which are the even- and
odd- eigenstates of the Hamiltonian with the nonlinear interaction term $%
U_0\left| \psi \left( x\right) \right| ^2$. Also the Thomas-Fermi
approximation $\psi _{TF}=\left[ \left( \mu -V\left( x\right) \right)
/U_0\right] ^{1/2}$ for $V\left( x\right) <\mu $ is given by the dashed
lines.

Now we turn to examine the difference of level splitting obtained from the
instanton method and the numerical simulation for the double-well trap. In
Fig. 2, $\Delta \mu $ on a logarithmic scale to base 10 as a function of the
chemical potential is depicted for barrier heights, $V_0=4$ and $V_0=5$ .
Results from both the numerical simulation (solid line) and instanton
approach (dotted line) exhibit an enhancement of the tunneling with
increasing chemical potential. To show how large the difference is, in the
insets, we plot the splitting divided by the exponential factor of the
analytical result. The results of the two approaches always have the same
order of magnitude and are close to each other. The periodic instanton
method evaluates fairly well the tunneling splitting even if the nonlinear
interaction is included. Quantitatively, however, it always over-estimates
the splitting for the whole range of the chemical potential. These can be
seen more clearly from the interacting constant $U_0$ dependence of the
level splitting displayed in Fig. 3 and its inset for $V_0=5$. We emphasize
here that a peculiar feature of this periodic instanton result for the level
splitting is that the prefactor depends on the chemical potential as
displayed in the insets of Fig. 2. This result is important, as has been
shown in \cite{Zhang}.

For the case of the optical lattice potential we still choose the same
parameters as in the case of the double-well. The nonlinear interaction
constant between atoms in the same well, $U_0=0.06$ for the repulsive
interaction. The typical profile of the condensate wave function $\psi
_e\left( x\right) $, shown by the solid line, and $\psi _o\left( x\right) $,
shown by the dotted lines, are plotted in Fig. 4 for $V_0=5$ and 150 atoms
in each well. The even wave function is symmetric, and the odd wave function
is maximally antisymmetric, i.e., the wavefunction segment in each well is
antisymmetric with respect to those of its neighbors. Fig. 5a and 5b display
the chemical potential dependence of the level splitting for $V_0=5$ and $%
V_0=4$, as determined by the method outlined above. Again, we find the
results from the periodic instanton method and GPE have the same order of
magnitude.

By comparing Fig. 2 and Fig. 5, it is shown that when the nonlinear
interaction between atoms is included, the level splitting is smaller than
the instanton result in the double-well case, but the splitting is larger in
the optical lattice case. This distinction depends on the structure of the
trapping potentials, for example, atoms in the optical lattice can tunnel
through the barrier in two directions, while the tunnel path for those in
the double-well is one-way only. This makes the periodic instanton result
different from the double well case. In spite of this, the periodic
instanton method remains good enough to describe the level splitting for the
BEC.

Recently a single bosonic Josephson junction \cite{Michael} has been
implemented by two weakly linked BEC in a double-well potential. In
previously reported realizations of condensates in double-well potentials 
\cite{Shin2} the time scale of tunneling dynamics was in the range of
thousands of seconds. In contrast, their setup allows the realization of
nonlinear tunneling times on the order of $50$ms, which makes the direct
observation of the nonlinear dynamics in a single bosonic Josephson junction
possible for the first time. We emphasize here the distinction between
tunneling in 3D and that in quasi-1D systems. The important parameter, the
``tunneling matrix element'' $K$ \cite{Smerzi} between two condensates is
related to the energy level splitting through $K=\Delta \mu /2$. The period
of oscillation $T$ can be obtained by numerically integrating eq. (1) in
Ref. \cite{Michael} for each given parameter $\Lambda =NU_0/2K$. In the 3D
case, $K$ is often assumed to be of the order $0.1$nK or $25$Hz, which gives
a relative small value for $\Lambda \sim 10$. In contrast, in the 1D case,
the tight confinement in the other 2 directions would suppress drastically
the tunneling and make the link between the condensates even weaker.
According to our calculation the parameter $K$ obviously takes typical
values of $10^{-3}\sim 1$Hz and $\Lambda $ may be as large as $10^3\sim 10^6$%
. As a consequence, the atoms tend to be trapped in the potential wells and
the observation of Josephson oscillation becomes almost impossible, e.g.,
the initial population imbalance must be less than $0.06$ for $\Lambda \sim
10^3$. A simple calculation shows that for $N=600,$ $U_0=0.06,$ $V_0=5$, the
Josephson oscillation may be observed with a period $2$ms, which is less
than that in the 3D experiments.

According to our calculation, the energy splitting is very small compared
with the chemical potential of the system and increases exponentially with
the chemical potential. The tunneling effect gives rise to the macroscopic
phase coherence of BEC across the barriers, which results in the significant
observable interference phenomena between different BEC segments.

\section{Summary}

In this paper we have investigated the tunnel splitting of the ground state
for a weakly-linked BEC trapped in double-well potential and in optical
lattice by solving the Gross-Pitaevskii equations numerically. It turns out
that the periodic instanton method is a reliable tool in the evaluation of
the tunnel splitting for the BEC. The reason for this is that for the
quantum tunneling\ through the potential barrier, the nonlinear interaction
is negligibly small and contributes overall to a finite chemical potential.
Our numerical scheme could easily be improved for the investigation of the
dynamical behavior of the condensates in multi-well potentials.

\begin{acknowledgments}
The authors acknowledge the NSF of China (Grant Nos. 10175039 and 90203007)
for financial support. YZ is also supported by Academy of Finland (Project
No 206108). This work was inspired by a poster by D. Diakonov, H. Smith and
C. Pethick. Y.J. Hao also thanks Y. Nie, L. Li and G. Zheng for the useful
discussions.
\end{acknowledgments}

Figure Captions:

1. Level splitting and wavefunctions of condensates confined in a double
well potential for $V_0=5.$ Panel (a) The symmetric (full line) and
antisymmetric (dotted lines) wave functions of the ground state with $U_0=0$
(noninteracting case). Panel (b) The same as Panel (a) but for $U_0=0.06$.
The Thomas--Fermi solution $\psi _{TF}=\left[ \left( \mu -V\left( x\right)
\right) /U_0\right] ^{1/2}$ is shown by the dashed lines.

2. Level splitting as a function of chemical potential $\mu $ for two values
of the barrier height $V_{0}=5$ (a) and $V_{0}=4$ (b). Insets: The splitting
divided by the exponential factor of the analytical result in the insets.
Solid lines: GPE results $\Delta \mu _{\text{GPE}}/\exp (-W)$, dotted lines:
periodic instanton results $\Delta \mu _{\text{Instanton}}/\exp (-W)$.

3. The interaction constant $U_0$ dependence of level splitting for
condensates in a double-well for $V_0=5$. The inset shows the splitting
divided by the exponential factor of analytical result. Solid lines: GPE
results, $\Delta \mu _{\text{GPE}}/\exp (-W)$, dotted lines: periodic
instanton results $\Delta \mu _{\text{Instanton}}/\exp (-W)$.

4. The symmetric (full line) and antisymmetric (dotted lines) wave functions
in an optical lattice for $V_0=5$.

5. Level splitting as a function of chemical potential $\mu $ in an optical
lattice for $V_0=5$ (a) and $V_0=4$ (b). The insets show the splitting
divided by the exponential factor of the analytical result. Solid lines: GPE
results $\Delta \mu _{\text{GPE}}/\exp (-W)$, dotted lines: periodic
instanton results $\Delta \mu _{\text{Instanton}}/\exp (-W)$.

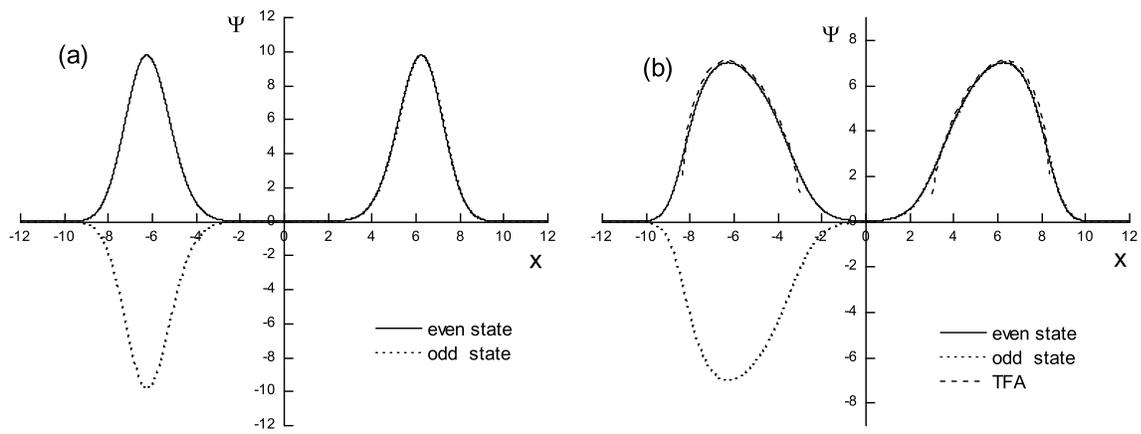
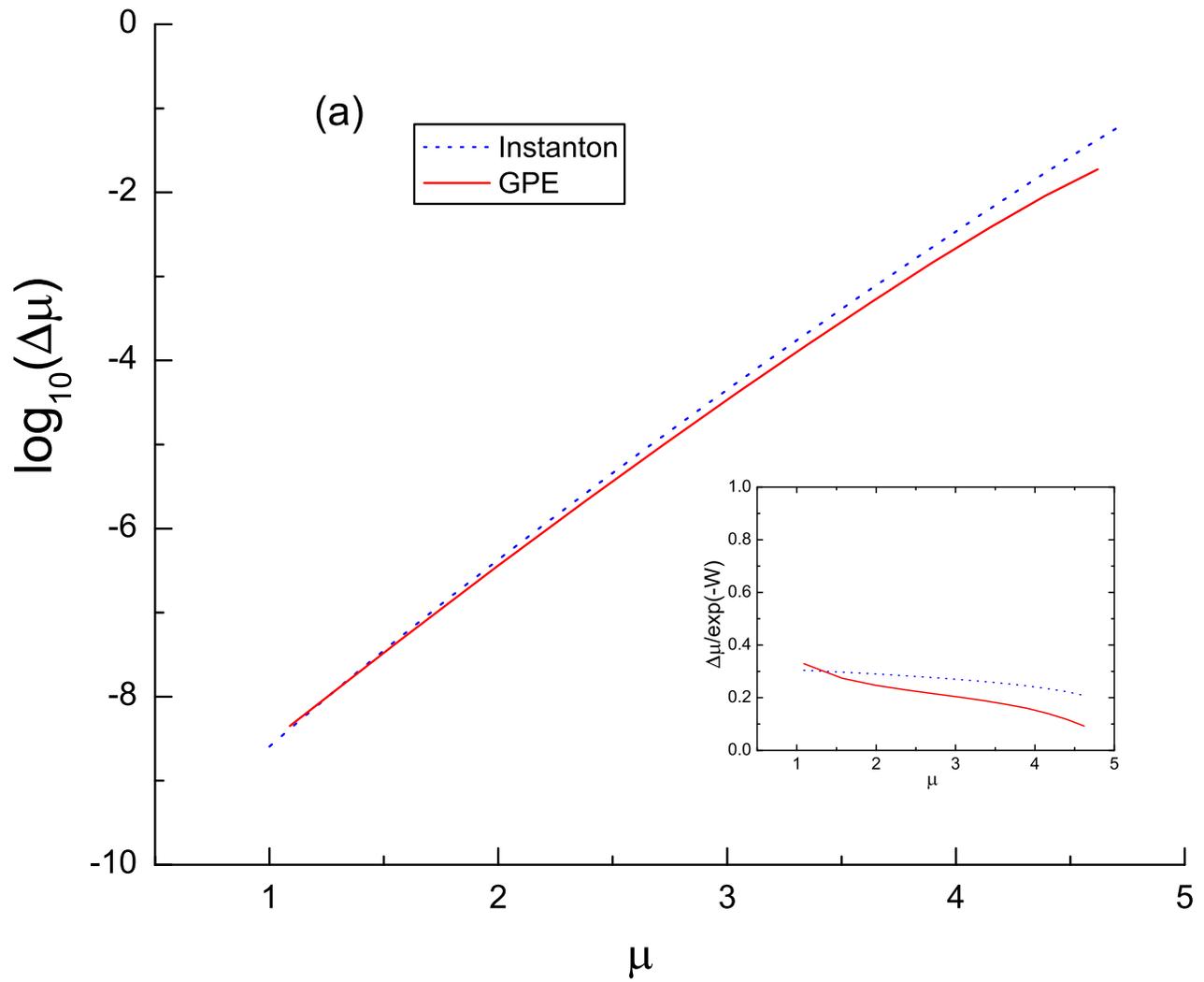
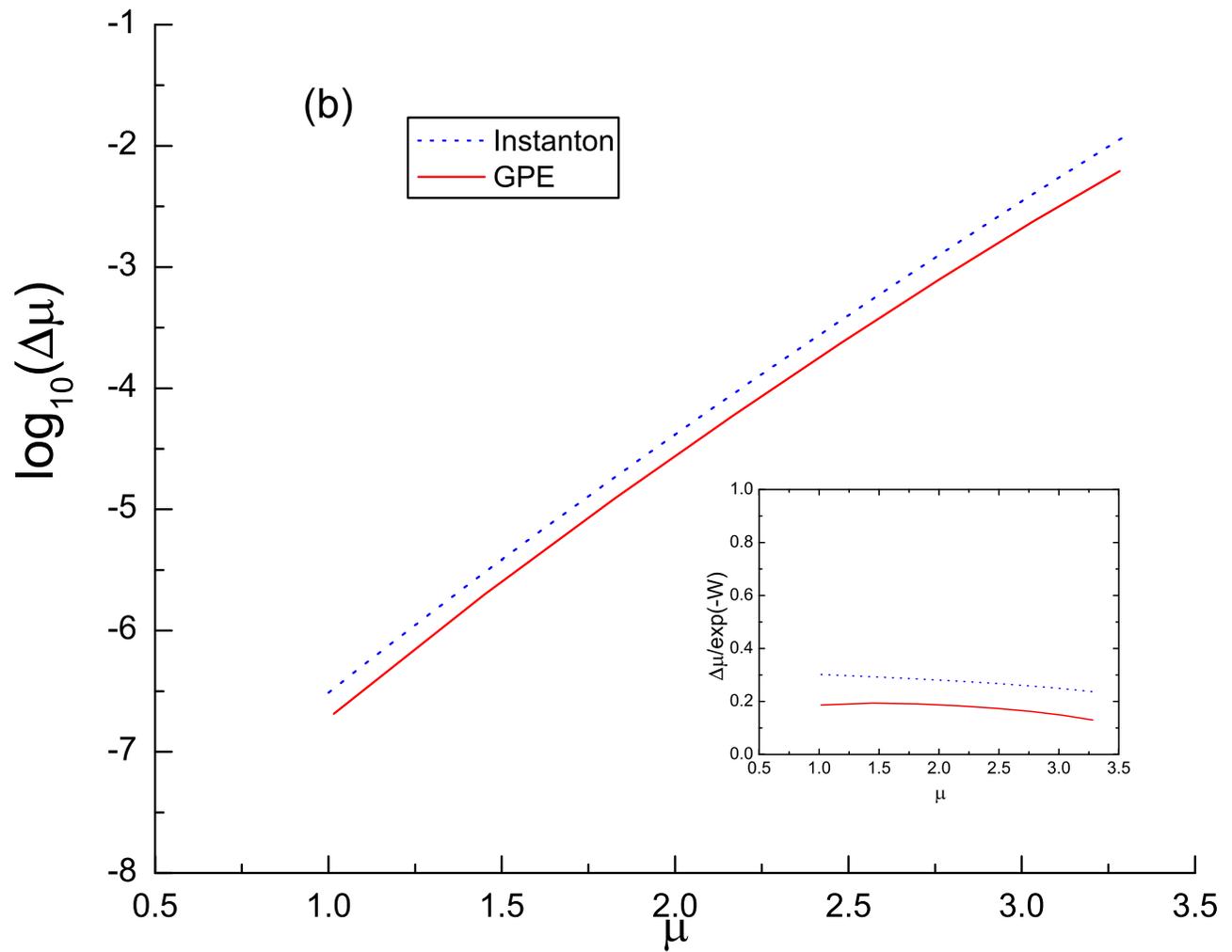
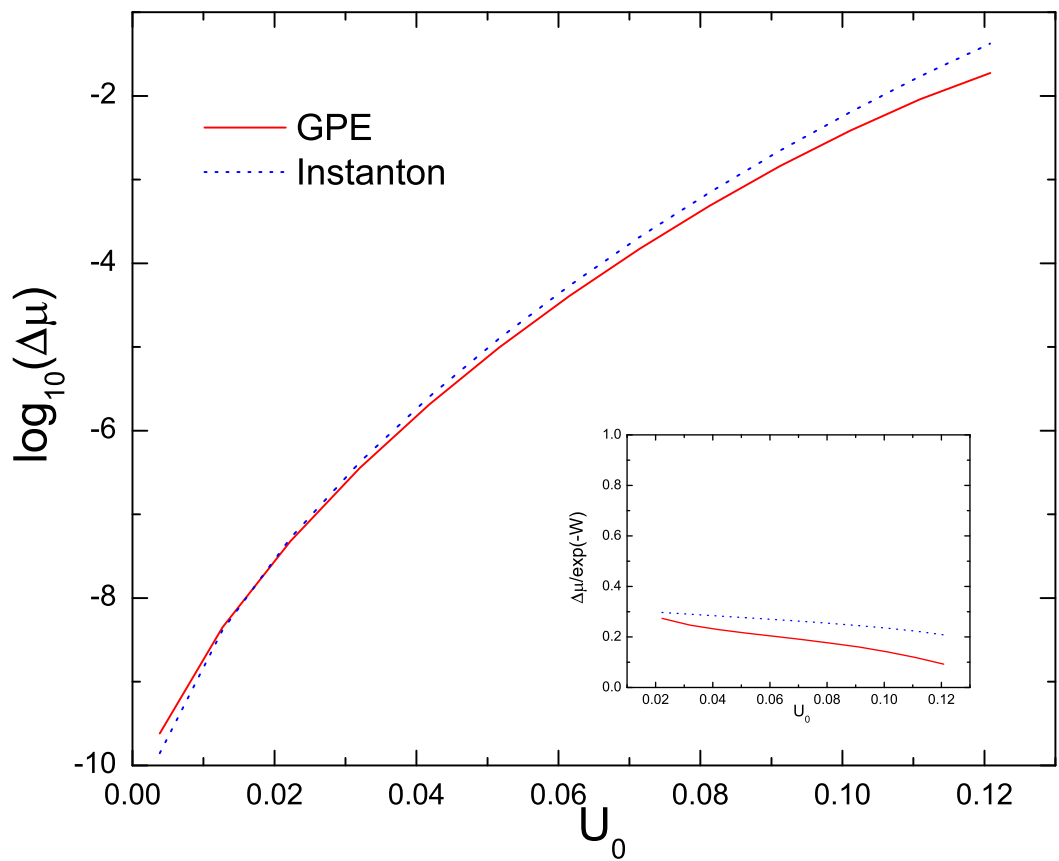
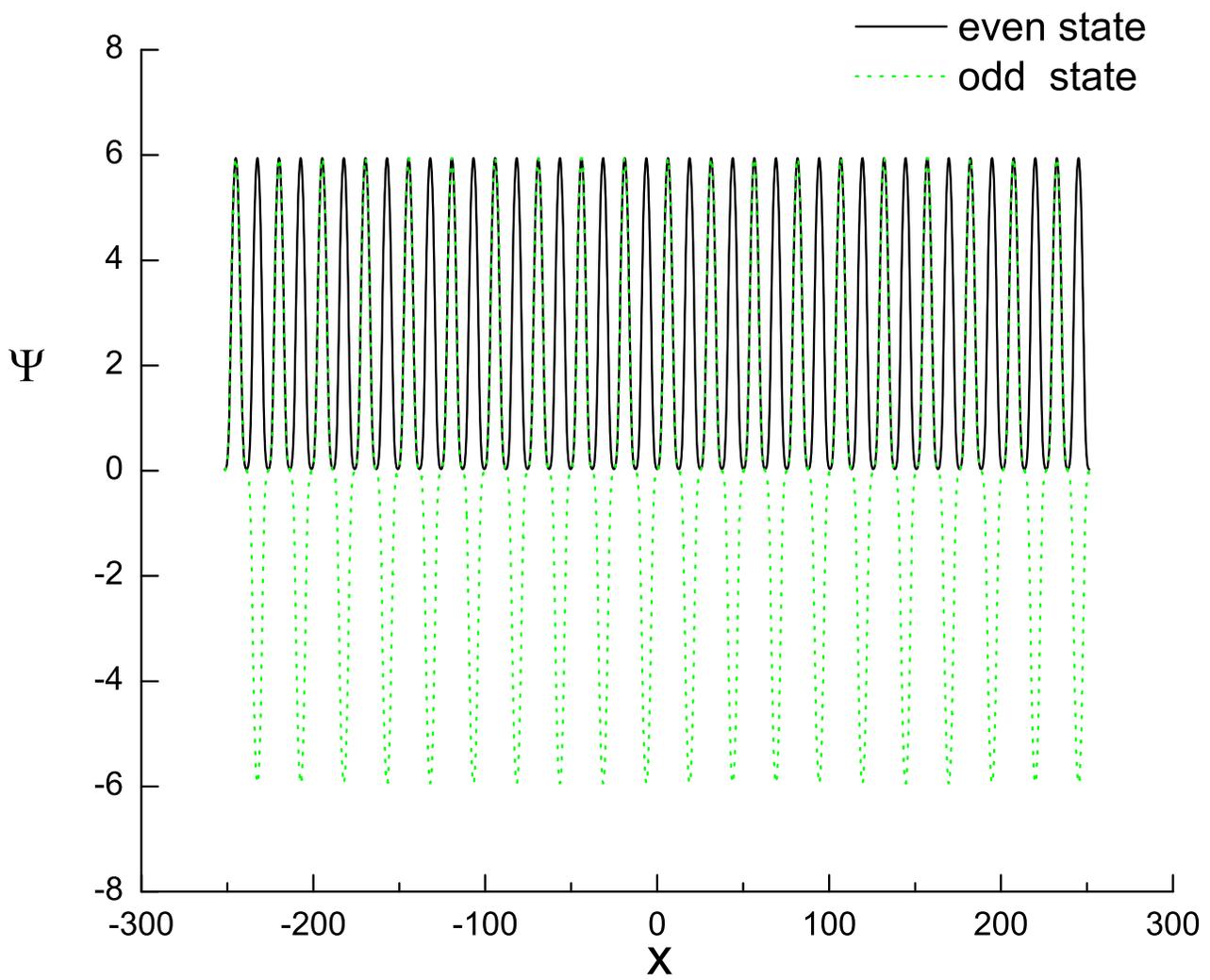
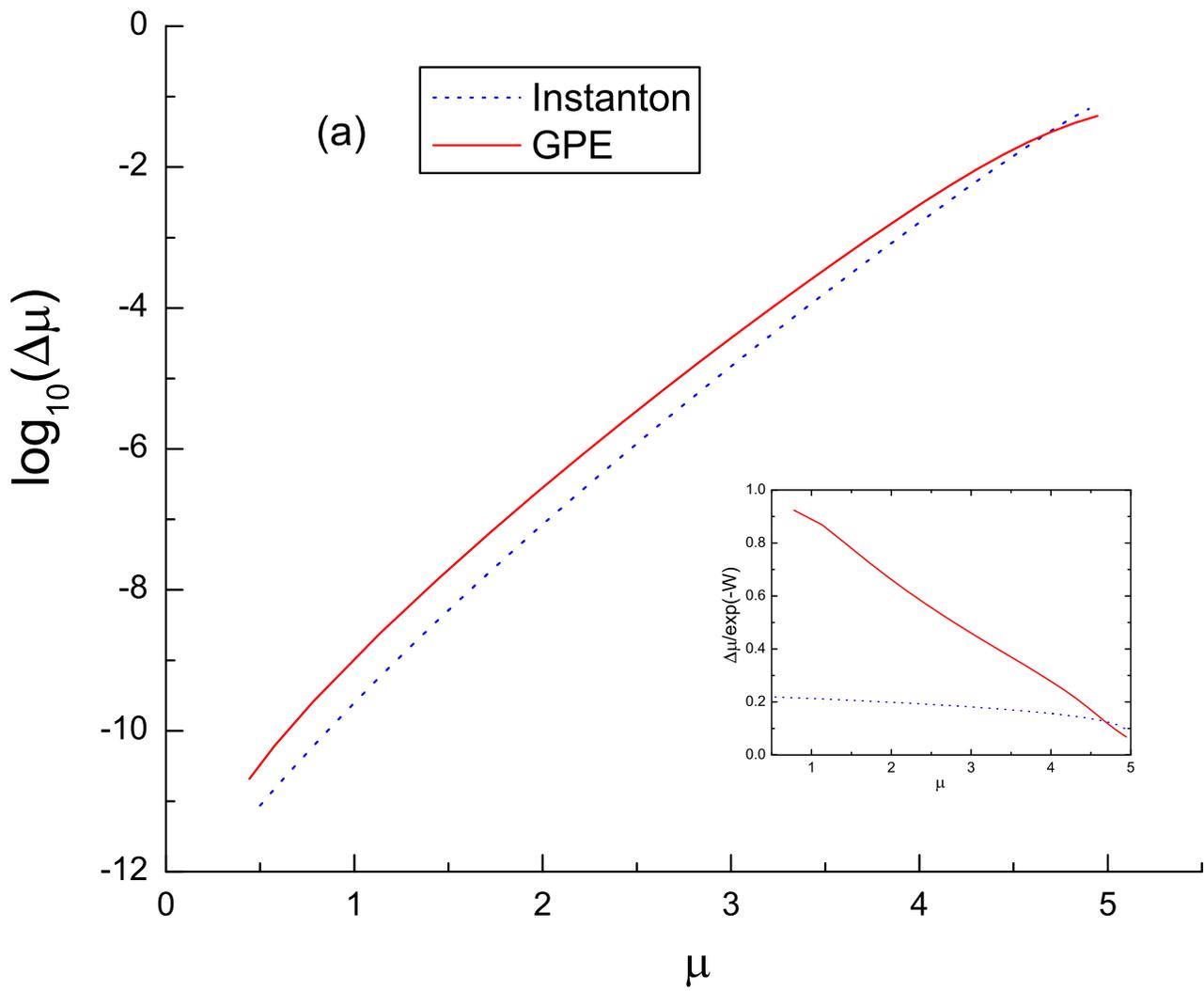
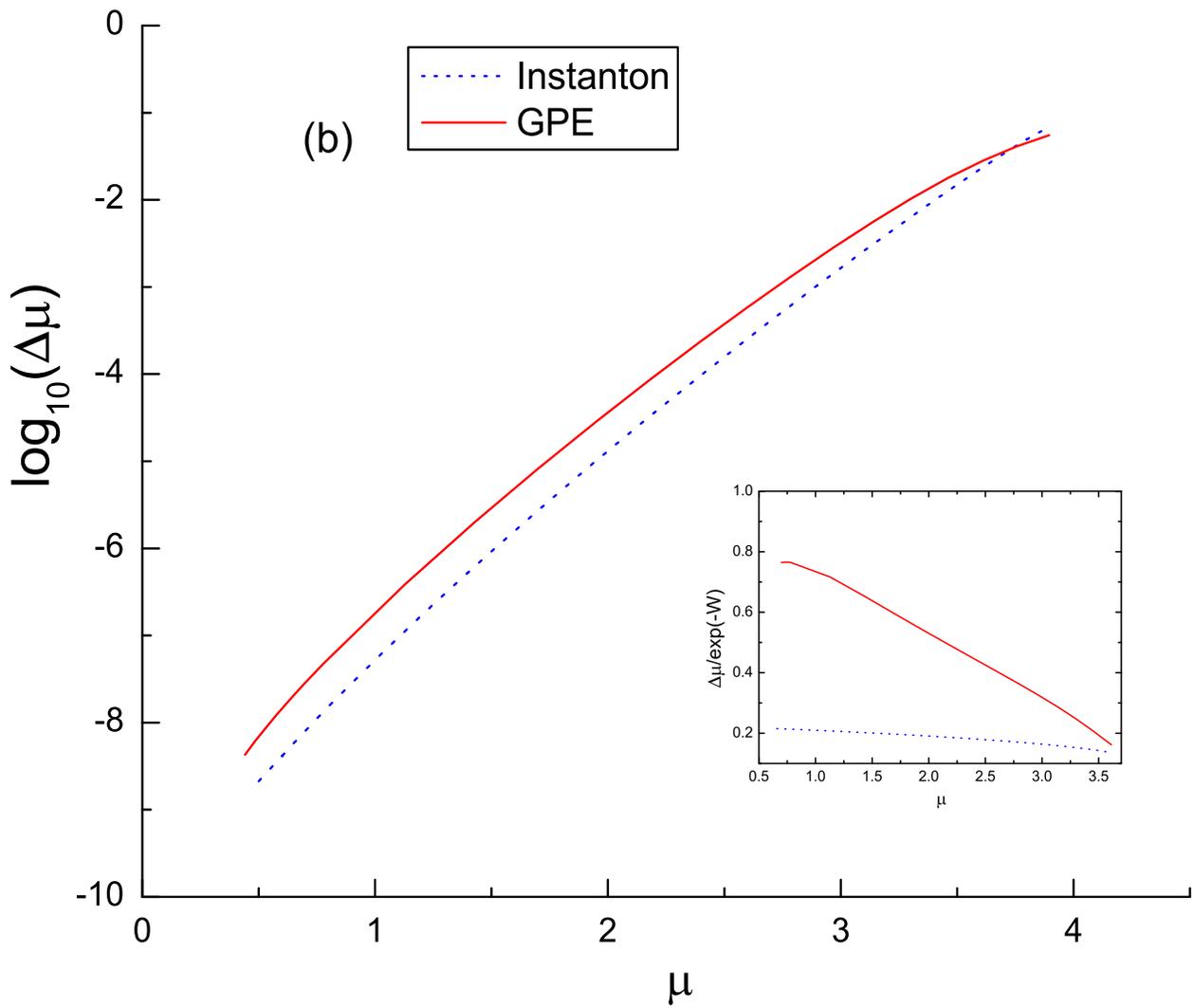
\end{document}